\documentstyle[12pt]{article}
\newskip\humongous \humongous=0pt plus 1000pt minus 1000pt

\newif\ifdtup

\def\ltap{\;\raisebox{-.4ex}{\rlap{$\sim$}} \raisebox{.4ex}{$<$}\;}
\def\gtap{\;\raisebox{-.4ex}{\rlap{$\sim$}} \raisebox{.4ex}{$>$}\;}

\def\beq{\begin{equation}}
\def\eeq{\end{equation}}

\def\beqn{\begin{eqnarray}}
\def\eeqn{\end{eqnarray}}
\relax

\def\dotx{\dotx{\dot\overline{x}}}

\relax
\catcode`\@=11

\def\@normalsize{\@setsize\normalsize{15pt}\xiipt\@xiipt
\abovedisplayskip 14pt plus3pt minus3pt%
\belowdisplayskip \abovedisplayskip
\abovedisplayshortskip  \z@ plus3pt%
\belowdisplayshortskip  7pt plus3.5pt minus0pt}

\def\small{\@setsize\small{13.6pt}\xipt\@xipt
\abovedisplayskip 13pt plus3pt minus3pt%
\belowdisplayskip \abovedisplayskip
\abovedisplayshortskip  \z@ plus3pt%
\belowdisplayshortskip  7pt plus3.5pt minus0pt

\def\@listi{\parsep 4.5pt plus 2pt minus 1pt
            \itemsep \parsep
            \topsep 9pt plus 3pt minus 3pt}}

\def\underline#1{\relax\ifmmode\@@underline#1\else
	$\@@underline{\hbox{#1}}$\relax\fi}
\@twosidetrue

\catcode`@=12

\catcode`\@=11

\catcode`\@=12

\relax

\def\figcap{\section*{Figure Captions\markboth
	{FIGURECAPTIONS}{FIGURECAPTIONS}}\list
	{Fig. \arabic{enumi}:\hfill}{\settowidth\labelwidth{Fig. 999:}
	\leftmargin\labelwidth
	\advance\leftmargin\labelsep\usecounter{enumi}}}
 \relax
\def\tablecap{\section*{Table Captions\markboth
	{TABLECAPTIONS}{TABLECAPTIONS}}\list
	{Table \arabic{enumi}:\hfill}{\settowidth\labelwidth{Table 999:}
	\leftmargin\labelwidth
	\advance\leftmargin\labelsep\usecounter{enumi}}}
 \relax
\def\reflist{\section*{References\markboth
	{REFLIST}{REFLIST}}\list
	{[\arabic{enumi}]\hfill}{\settowidth\labelwidth{[999]}
	\leftmargin\labelwidth
	\advance\leftmargin\labelsep\usecounter{enumi}}}
 \relax

\catcode`\@=11

\def\FERMIPUB{}
\def\FERMILABPub#1{\def\FERMIPUB{#1}}
\def\ps@headings{\def\@oddfoot{}\def\@evenfoot{}
\def\@oddhead{\hbox{}\hfill
	\makebox[.5\textwidth]{\raggedright\ignorespaces --\thepage{}--
	\hfill {\rm FERMILAB--Pub--\FERMIPUB}}}
\def\@evenhead{\@oddhead}
\def\subsectionmark##1{\markboth{##1}{}}
}

\ps@headings

\relax
\FERMILABPub{92/171--T}
\begin{document}
\begin{titlepage}
\def\ba{\begin{array}}
\def\ea{\end{array}}
\def\thefootnote{\fnsymbol{footnote}}
\begin{flushright}
	FERMILAB--PUB--92/171--T\\
	June 1992
\end{flushright}
\vfill
\begin{center}
{\large \bf Structure-Function Approach to Vector-Boson Scattering in
pp Collisions} \\
\vfill
	{\bf T.~Han, G.~Valencia, and S.~Willenbrock}\footnotemark[1]\\
\footnotetext{[*]Permanent address: Physics Department, Brookhaven
National Laboratory, Upton, NY
11973}
	{\it Fermi National Accelerator Laboratory \\
	P.O. Box 500, Batavia, IL 60510}
\vfill
\end{center}
\begin{abstract}
We discuss weak-vector-boson scattering, at
next-to-leading order in QCD, within the framework
of hadronic structure functions.
We use this approach to calculate the Higgs-boson
production cross section via vector-boson fusion at the LHC/SSC;
we find a modest increase over the leading-order
prediction. We also give expressions for the distribution of vector
bosons in a proton (effective-$W$ approximation) including ${\cal O}
(\alpha_s)$ corrections.
\end{abstract}

The LHC/SSC will be the first machines capable of studying
weak-vector-boson scattering, where the ``initial'' vector bosons are
radiated from the quarks and antiquarks which reside in the proton. It
is believed that this process will yield clues to the mechanism which
breaks the electroweak symmetry \cite{CG}, which is one of the outstanding
puzzles of high-energy physics. For example, in the standard Higgs
model of electroweak-symmetry breaking,
the Higgs boson is copiously produced
via vector-boson fusion \cite{CD,KRR}.

In this paper we show that weak-vector-boson scattering may be
calculated in terms of the charged-current and neutral-current
hadronic structure functions $F_i(x,Q^2)$ $(i=1,2,3).$
One advantage of
this approach is that the ${\cal O}(\alpha_s)$ QCD correction to
vector-boson scattering can be incorporated simply by employing the
QCD-corrected expressions for the structure functions in the parton
model. This constitutes a considerable simplification of the calculation.
The resulting cross section is differential in the vector-boson
scattering subprocess. A similar
approach has been applied to two-photon processes \cite{CMT}.

As usual, we write the hadronic tensor
$W_{\mu\nu}$ in terms of the three structure functions:
\beqn
MW_{\mu\nu}(x,Q^2) & = & F_1(x,Q^2) \bigg (-g_{\mu\nu} +
\frac{q_\mu q_\nu}{q^2}\bigg)\nonumber\\
& + & \frac{F_2(x,Q^2)}{P{\cdot} q}\bigg(P_\mu - \frac{P{\cdot} q}{q^2}
q_\mu\bigg)\bigg(P_\nu-\frac{P{\cdot} q}{q^2}q_\nu\bigg)
\nonumber\\
& - & i \frac{F_3(x,Q^2)}{2P{\cdot} q} \epsilon _{\mu\nu\rho\sigma} P^\rho
q^\sigma
\label{ca}
\eeqn
where $M$ is the proton mass,
$P_\mu$ is the proton four-momentum, $q_\mu$ is the
vector-boson four-momentum, $Q^2=-q^2,$ and $x=Q^2/2 P{\cdot} q$. Vector-boson
scattering in $pp$ collisions is then calculated by contracting the
hadronic tensors at each vertex with the tensor corresponding to the
square of the vector-boson-scattering subprocess, as in Fig.~1. We find
\beqn
d\sigma & = & \frac{1}{2S} 4\frac{g_{V_1}^2}{8}\frac{g_{V_2}^2}{8}
\frac{1}{(Q_1^2+M^2_{V_1})^2}\frac{1}
{(Q^2_2+
M^2_{V_2})^2}M^2W_{\mu\nu}(x_1, Q^2_1){\cal M}^{\mu\rho}
{\cal M}^{*\nu\sigma}
W_{\rho\sigma}(x_2,Q^2_2)\nonumber\\
& &
\frac{d^3P_{X_1}}{(2\pi)^32E_{X_1}}\frac{d^3P_{X_2}}{(2\pi)^32E_{X_2}}
ds_1
ds_2 d\Gamma (2\pi)^4\delta^4(P_1+P_2-P_{X_1}-P_{X_2}-P_{VV})
\label{cd}
\eeqn
where $g^{}_W=g,g^{}_Z=g/\cos\theta_W$, $S$ is the square of the total
machine energy,
$s_i=(P_i+q_i)^2$ is the squared invariant
mass of the remnant of proton $i$, and $d\Gamma$ is the vector-boson-scattering
phase space.

At lowest order,
no color is exchanged between the protons, and the remnants of the protons
are color singlets.
It follows that no QCD corrections due to gluon exchange
between the first proton (or its remnant) and the second proton (or its
remnant), or between either proton (or their remnants)
and the products of the vector-boson-scattering subprocess,
occur at ${\cal O}(\alpha_s)$.
Therefore, the ${\cal O}(\alpha_s)$
correction to vector-boson scattering may be factorized into the corrections to
the structure functions and the corrections to the vector-boson-scattering
subprocess. This factorization is no longer true at ${\cal O}(\alpha_s^2)$,
however.
The structure functions are given at
next-to-leading order in QCD by \cite{AEM},
\beqn
F_1(x,Q^2) & = & \sum_i\bigg(C^2_{Vi} +C^2_{Ai}\bigg) \bigg\{ \int^1_x
\frac{dy}{y} \bigg[q_i(y,Q^2)+\bar q_i (y, Q^2)\bigg]\nonumber\\
& & \bigg[ \delta (1-z) -\frac{4}{3} \frac{\alpha_s(Q^2)}{\pi} z\bigg]
-2\frac{\alpha_s(Q^2)}{\pi}
\int^1_x \frac{dy}{y} g(y, Q^2) z(1-z) \bigg\}\nonumber\\
F_2(x,Q^2)& = & 2x\sum_i\bigg(C^2_{Vi} +C^2_{Ai}\bigg)
\bigg[q_i(x,Q^2)+\bar q_i (x, Q^2)\bigg]\nonumber\\
F_3(x,Q^2) & = & 4 \sum_i C_{Vi}C_{Ai} \bigg\{
\int^1_x \frac{dy}{y}\bigg[-q_i(y, Q^2)+\bar q_i(y,Q^2)\bigg]\nonumber\\
& & \bigg[\delta(1-z) - \frac{2}{3}\frac{\alpha_s(Q^2)}{\pi}(1+z)\bigg]
\bigg\}
\label{cb}
\eeqn
where $z=x/y$,
the sum runs over the flavors of all quarks and antiquarks which
contribute to a given structure function,
$C_{Vi} = C_{Ai} = 1/\sqrt 2$ for $W^\pm$, and
$C_{Vi} = T^3_{Li}-2e_i\sin^2\theta_W$, $C_{Ai} = T^3_{Li} =\pm \frac{1}{2}$
for $Z$. These expressions
correspond to the DIS factorization scheme.  The corresponding
expressions in the $\overline{\rm MS}$ scheme are obtained by
replacing the
quark and antiquark distribution functions in the leading-order terms
by \cite{BBDM}
\beqn
q_i(x,Q^2) & = & q_i^{\overline{MS}}(x,Q^2)\nonumber\\
& + & \frac{1}{4} \frac{\alpha_s(Q^2)}{\pi} \int^1_x \frac{dy}{y} g(y, Q^2)
\bigg[[z^2+(1-z)^2]\ln\bigg(\frac{1-z}{z}\bigg)+8z(1-z)-1\bigg]\nonumber\\
& + & \frac{2}{3}\frac{\alpha_s(Q^2)}{\pi} \int^1_x
\frac{dy}{y}q_i(y,Q^2)\bigg[(1+z^2)\bigg(\frac{\ln(1-z)}{1-z}\bigg)_{+}
-\frac{3}{2}\frac{1}{(1-z)_{+}}\nonumber\\
& - & \frac{1+z^2}{1-z}\ln z
+3+2z-(\frac{9}{2}+\frac{\pi^2}{3})\delta(1-z)\bigg]
\label{ms}
\eeqn
where the ``plus'' prescription is defined as usual (see, e.~g.,
Ref.~\cite{AEM}).

If the structure functions, over the relevant ranges of $x$ and $Q^2$,
were available from deep-inelastic scattering, they could be
used directly, bypassing the parton model altogether. The typical
$Q^2$ in vector-boson-scattering processes is $M^2_V,$ which is
within the reach of HERA, but only for
$x\gtap 0.1$ \cite{HERA}.
The possibility of using the measured structure functions directly will thus
be limited to very high invariant masses for the
vector-boson-scattering subprocess, $M_{VV}\sim 0.1\; \sqrt S$.
LEP $\times$ LHC would be able to reach down to $x\sim 0.01$ at this value of
$Q^2$ \cite{LHC}.

The validity of this approach relies on the factorization of the cross
section that is inherent in
Fig.~1.  In order for the factorization to hold, two criteria must be
satisfied.
First, there must be no significant
interference, at the parton level, between diagrams in which a quark
from one proton  and a quark from the other proton scatter into the same
final-state quark.
Kinematic arguments suggest that this
interference is very small \cite{DW}, and we have verified
that this is true in the
example below. Second, the vector-boson-subprocess final state must
be produced only via this mechanism, or dominantly so, or not
interfere with similar final states.  Examples include Higgs-boson
production \cite{CD,KRR}, heavy-fermion production \cite{DW2}, and terms
of enhanced electroweak strength in longitudinal-vector-boson scattering
\cite{CG,KS}.

As an explicit example of this
approach, we calculate Higgs-boson production via vector-boson fusion
at the LHC/SSC. The differential cross section is given by Eq.~(\ref{cd}), with
\beqn
\lefteqn{M^2W_{\mu\nu}(x_1, Q^2_1){\cal M}^{\mu\rho}{\cal M}^{*\nu\sigma}
W_{\rho\sigma}
(x_2,Q^2_2) = g_V^2 M_V^2 \bigg\{
F_1 (x_1, Q^2_1) F_1 (x_2, Q_2^2)\bigg[2 +
\frac{(q_1{\cdot} q_2)^2}{q_1^2q_2^2}\bigg] }\nonumber \\
&& +  {F_1(x_1, Q_1^2) F_2(x_2, Q_2^2)\over P_2 {\cdot}
q_2}\bigg[\frac{(P_2{\cdot} q_2)^2}{q^2_2} -
M^2+\frac{1}{q_1^2}\bigg(P_2{\cdot} q_1-\frac{P_2{\cdot} q_2}{q^2_2}
q_1{\cdot} q_2\bigg)^2\bigg]\nonumber\\
&& +  {F_2 (x_1, Q^2_1) F_1(x_2, Q^2_2)\over P_1{\cdot}
q_1} \bigg[\frac{(P_1{\cdot} q_1)^2}{q_1^2}-M^2 +
\frac{1}{q_2^2}\bigg(P_1{\cdot} q_2 - \frac{P_1{\cdot} q_1}{q_1^2}
q_1{\cdot} q_2\bigg)^2\bigg]\nonumber\\
&& + {F_2(x_1, Q_1^2) F_2(x_2, Q^2_2)\over P_1{\cdot}
q_1 \ P_2{\cdot} q_2}
\bigg(P_1{\cdot} P_2 - \frac{P_1{\cdot} q_1 P_2{\cdot} q_1}{q_1^2}-
\frac{P_2{\cdot} q_2P_1{\cdot} q_2}{q^2_2}
+ \frac{P_1{\cdot} q_1 P_2{\cdot}
q_2  q_1{\cdot} q_2}{q_1^2 q^2_2}\bigg)^2\nonumber\\
&& + {F_3 (x_1, Q_1^2) F_3 (x_2, Q^2_2) \over 2\
P_1{\cdot} q_1 \ P_2{\cdot} q_2}
\bigg( P_1{\cdot} P_2 q_1{\cdot} q_2 -P_1{\cdot} q_2P_2 {\cdot} q_1\bigg)
\bigg\}.
\label{ce}
\eeqn
For $W^+W^-$ fusion one must sum over
the $W^+$ (and the $W^-$) being emitted from either proton.

We show in Fig.~2 the total cross section for Higgs-boson production
at the
LHC/SSC through ${\cal O}(\alpha_s)$, using the next-to-leading-order
parton distribution functions of Ref.~\cite{MT}, set S1-DIS.
Also shown are the leading-order cross sections, evaluated
using the leading-order parton distribution functions of Ref.~\cite{MT},
which were fit to the same data as set S1-DIS.
The next-to-leading-order cross section at the SSC is $12\%$ larger at
$m_H=$100 GeV,
and $6\%$ larger at $m_H=$ 800 GeV, than the leading-order cross section.
The corresponding increase at the LHC is $8\%$ and $6\%$.
Most of the increase is due to the parton distribution functions,
not to the explicit ${\cal O}(\alpha_s)$ corrections to the
structure functions, because the total cross section is dominated by
$F_2$, which receives no QCD correction in the DIS scheme.
To the extent that the next-to-leading-order
parton distribution functions provide a better fit to the available
data, and a more reliable extrapolation to high $Q^2$,
the next-to-leading-order cross sections in Fig.~2 represent the
best predictions
possible at this time.

The structure-function approach to vector-boson scattering makes it
clear that the relevant scale of the structure functions, and
therefore of the parton distribution functions, is the $Q^2$ of the
vector boson. Since the typical $Q^2$ is about $M^2_V$, the use of this
fixed scale is a good approximation.
Since the parton model
breaks down at low $Q^2$, and the parton distribution functions
are available only for $Q^2 > 4 \;{\rm GeV}^2$, we have not included
the region $Q^2 < 4\; {\rm GeV}^2$ in our calculation. We estimate
that this region contributes only about
$4\; {\rm GeV}^2/M_V^2\sim 10^{-3}$ of the total cross section.

A heavy Higgs boson
decays predominantly to $W$- and $Z$-boson pairs. The principal
background to this signal at the LHC/SSC is $W^+W^-$ and $ZZ$
production via quark-antiquark annihilation. The QCD correction to
the $ZZ$ invariant-mass distribution increases the leading-order prediction
by about $25\%$ near threshold, up to about $60\%$
at 800 GeV (for a factorization scale $\mu^2=M^2_{ZZ}$)
\cite{OO}.
The corresponding increase for $W^+W^-$ is
about $45\%$ near threshold and about $75\%$ at 800 GeV \cite{O}.
The absence of a comparable increase in the
Higgs-boson production cross section therefore decreases the
signal-to-background ratio.

In the standard Higgs model, the Higgs
boson is also produced at the LHC/SSC from gluon fusion via a
top-quark loop \cite{GGMN} with a comparable or larger cross section
than from vector-boson fusion
(for $m_t >$ 100 GeV, $m_H <$ 800 GeV). The QCD
correction to this process is known only for a light $(m_H < 2m_t)$ Higgs boson
\cite{D}, and increases the cross section by $50\%$ at the SSC and
$100\%$ at the LHC (for $\mu^2=m_H^2$).
It would be desirable to observe Higgs-boson
production via gluon fusion and vector-boson fusion separately, since the
former involves the coupling of the Higgs boson to the top quark, while
the latter does not.

If the invariant mass of the vector-boson-scattering subprocess is large
compared to $M_V$, the cross section is dominated by the region
$Q^2\ltap M_V^2$. One can then derive a
distribution function for vector bosons carrying a fraction
$x$ of the proton's momentum \cite{KRR,CG2}.  In terms
of hadronic structure functions, we find
\beqn
f_T (x) &= & {g_V^2 \over 32 \pi^2} \; \frac{1}{x}\;\int^1_x
\;\frac{dy}{y} \;\left[F_2 \left(y,M^2_V\right)
\left(1-\frac{x}{y}\right) + F_1 \left(y,M^2_V\right)
\frac{x}{y}^2\right]\ln \bigg (1+\frac{Sy(y-x)}{M_V^2}\bigg) \nonumber \\
f_L (x) &= & {g_V^2 \over 32 \pi^2} \; \frac{1}{x}\;\int^1_x
\;\frac{dy}{y} \;\left[F_2 \left(y,M^2_V\right)
\left(1-\frac{x}{2y}\right)^2 -\frac{1}{2} F_1 \left(y,M^2_V\right)
\frac{x}{y}^2\right]
\label{cf}
\eeqn
where the subscripts $T,L$ denote the polarization of the vector
boson (the two transverse polarizations have been averaged).  At leading order
in QCD, $F_2 (y, Q^2 )=2y F_1 (y,Q^2)$ (Callan-Gross relation), and the
expressions
above reduce to the usual parton-model formulae.  The ${\cal O}(\alpha_s )$
expressions for the vector-boson distribution functions are obtained
simply by using the next-to-leading-order expressions for the structure
functions, Eq.~(\ref{cb}). It has already been shown that there are no terms of
${\cal O}(\alpha_s \pi)$ (from soft gluons)
in the DIS scheme \cite{D3} (although there are
in the $\overline{\rm MS}$ scheme), as is evident from Eq.~(\ref{cb}).
In any case, the ${\cal O}(\alpha_s \pi)$
terms generally do not dominate at LHC/SSC energies.

{\bf Acknowledgements}

We are grateful for conversations with S.~Dawson, D.~Dicus, K.~Ellis, W.~Giele,
J.~Morfin, C.~Quigg,
G.~J.~van Oldenborgh, and W.-K.~Tung.    T.~H. was
supported by an SSC Fellowship from the Texas National Research Laboratory
Commission (TNRLC) under Award No.~FCFY9116.  S.~W. was supported by an
SSC Fellowship from the TNRLC
and by contract number
DE-AC02-76-CH-00016 with the U.S. Department of Energy.
\vfill
\eject


\newpage
\begin{figcap}
\item Structure-function approach to vector-boson scattering in $pp$
collisions.
\item Total cross section for Higgs-boson production via vector-boson
fusion at the (a) LHC and (b) SSC, versus the Higgs-boson mass.
The solid curves are calculated at next-to-leading order in QCD, using set
S1-DIS of Ref.~\cite{MT}.  The dashed curves are calculated at leading order,
using the leading-order set of Ref.~\cite{MT}.  The cross sections due to
an intermediate $W^+W^-$ and $ZZ$ pair are shown separately.
\end{figcap}
\end{document}